\documentclass[12pt]{article} 
\usepackage{amsmath,amsthm,amssymb,bm,mathrsfs}
\usepackage{graphicx,subfigure}
\usepackage{caption,enumerate,listings,setspace}
\usepackage[colorlinks,linkcolor=red,anchorcolor=blue,citecolor=blue]{hyperref}
\usepackage[margin=1in]{geometry}
\usepackage[title]{appendix}
\usepackage{enumitem}
\theoremstyle{plain}
\usepackage{booktabs}
\usepackage[numbers]{natbib}
\usepackage{array}
\usepackage{algorithm}
\usepackage{algpseudocode}
\usepackage{makecell}
\usepackage{tikz}
\usepackage{bbding}
\usepackage{cancel}
\usepackage{amsthm}
\usepackage{graphicx}
\usepackage{wrapfig}
\usepackage{caption}
\theoremstyle{definition}

\numberwithin{example}{section}

\newtheorem{definition}{{\bf Definition}}

\usepackage[utf8x]{inputenc}
\usepackage{mathrsfs}
\usepackage{amssymb}
\usepackage{amsfonts}
\usepackage{amsmath}
\usepackage{amsthm,verbatim}
\usepackage{wrapfig}
\usepackage{multirow}
\usepackage{longtable}
\usepackage{enumerate}
\usepackage{color, graphicx}
\usepackage{tikz}

\def\boxit#1{\vbox{\hrule\hbox{\vrule\kern6pt
          \vbox{\kern6pt#1\kern6pt}\kern6pt\vrule}\hrule}}

\title{\textbf{\texttt{DEREC-SIMPRO}: unlock Language Model benefits \\ to advance Synthesis in Data Clean Room}} 

\author
{ Tung Sum Thomas Kwok \thanks{Master's Student,  Boston University, MA, 02134. Email: tk1018@bu.edu}, 
Chi-Hua Wang \thanks{Postdoctoral Scholar, Department of Statistics and Data Science,  UCLA, CA, 90095. Email: chihuawang@ucla.edu},
Guang Cheng\thanks{Professor, Department of Statistics and Data Science, UCLA, CA, 90095. Email: guangcheng@ucla.edu}
}

\begin{document} 

\maketitle

\begin{abstract}
Data collaboration via Data Clean Room offers value but raises privacy concerns, which can be addressed through synthetic data and multi-table synthesizers. Common multi-table synthesizers fail to perform when subjects occur repeatedly in both tables. This is an urgent yet unresolved problem, since having both tables with repeating subjects is common. To improve performance in this scenario, we present the \texttt{DEREC} 3-step pre-processing pipeline to generalize adaptability of multi-table synthesizers. We also introduce the \texttt{SIMPRO} 3-aspect evaluation metrics, which leverage conditional distribution and large-scale simultaneous hypothesis testing to provide comprehensive feedback on synthetic data fidelity at both column and table levels. Results show that using \texttt{DEREC} improves fidelity, and multi-table synthesizers outperform single-table counterparts in collaboration settings. Together, the \texttt{DEREC-SIMPRO} pipeline offers a robust solution for generalizing data collaboration, promoting a more efficient, data-driven society.
\end{abstract}

\bigskip
\noindent{\bf Key Words:} Data Collaboration, Data Clean Room, Large Language Models, Data Synthesis, Synthetic Data Evaluation

\begin{figure}
    \centering
    \includegraphics[width = 0.95\linewidth]{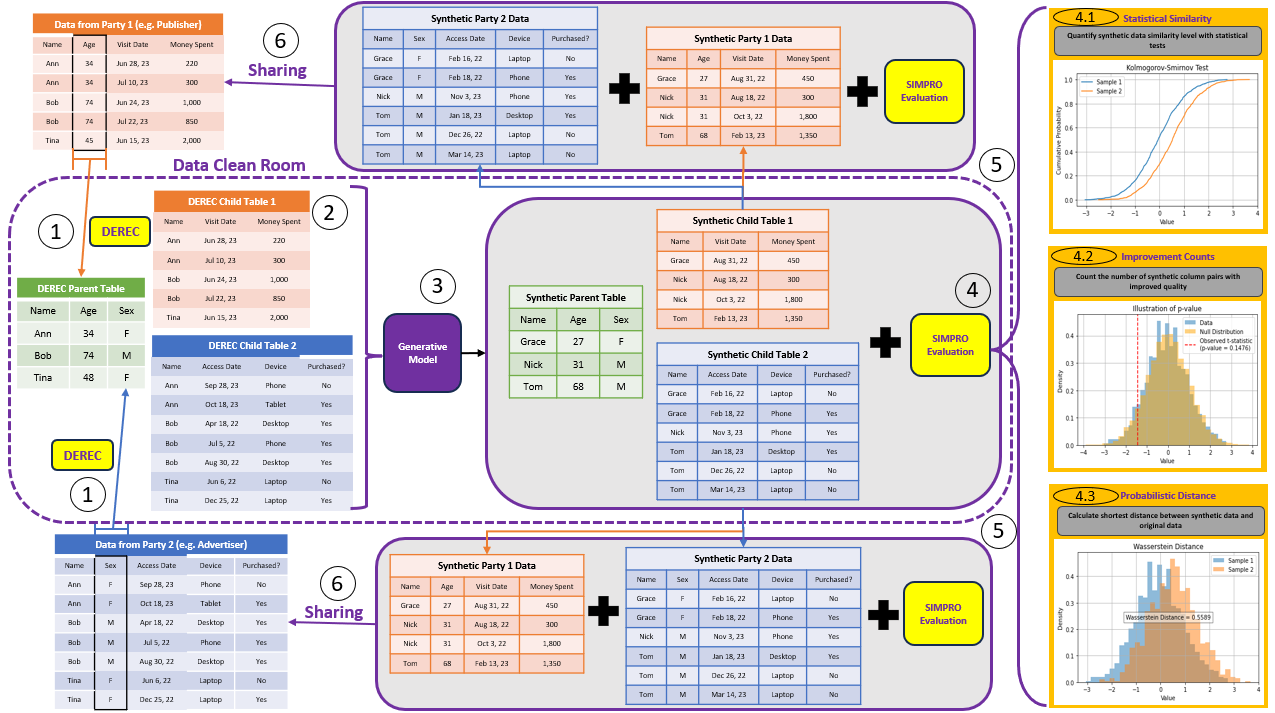}
    \caption{Proposed \texttt{DEREC} 3-steps pre-processing pipeline and \texttt{SIMPRO} 3-aspects synthetic data evaluation metrics. (1) Party 1 and Party 2 contribute their data in Data Clean Room (2) Construct parent and child columns with \texttt{DEREC} (Section \ref{sub:derec}) (3) Generate synthetic data with language model-based multi-table synthesizer (4) Audit the quality of synthetic data through \texttt{SIMPRO} evaluation metrics (Section \ref{sub:simpro}): (4.1) Compute statistical similarity between original and synthetic data (Section \ref{subsub:statistical-similarity}) (4.2 \& 4.3) Compare synthetic data quality generated by different models (Section \ref{sub:conditional_distribution_indicator}) using statistical equality test p-values (4.2) and distance metrics (4.3)   (5) Augment the first (second) party real data with synthetic second (first) party data (6) Send the augmented tabular data back to both parties to share the data}
    \label{fig:overview}
\end{figure}

\clearpage
\section{Introduction}
Data collaboration promotes innovation by leveraging the collective expertise of various stakeholders, e.g., two initiatives in Nepal have developed a platform to gather resources from different parties to improve local health and population data systems \cite{nepal}. Increasingly, enterprises are turning to Data Clean Rooms, which connect and synthesize datasets from multiple sources \cite{hbs-data-collaboration} \cite{ParraMoyano2024DataCollaboration}. Multi-table synthesizers address both data synthesis and integration needs, sparking interest in incorporating Language Model backbones for data generation, particularly following the success of ChatGPT \cite{gpt4o}.\\
\\
In managing a generation-based Data Clean Room that involves synthesis and evaluation processes, a key challenge for each is identified. During the synthesis process, existing multi-table synthesizers require a strict one-to-many table relationship, and any deviation from this can result in underperformance. However, real-life datasets often feature subjects that occur repeatedly (Section \ref{sub:cross-table_management}), which existing synthesizers are not equipped to handle due to their inability to manage many-to-many relationships. This limitation negatively impacts synthesizer performance (Refer to Section \ref{sub:improvement-counts-results}). In the evaluation step, current metrics are inefficient in assessing synthetic data quality in multi-table contexts, e.g., the \texttt{Synthetic Data Metrics} ("\texttt{SDMetrics}") \cite{sdmetrics} tends to favor 'smoothed' results without accurately reflecting shape similarity. \\ 
\\
The first challenge involves effectively establishing connections between two tables with many-to-many relationships, as the potential combinations for each column can be infinite unless additional information from the datasets helps differentiate specific observations. The second challenge focuses on defining an indicator that accurately measures the similarity between features of the original and synthetic datasets.\\
\\
In summary, the contributions of this work include:
\begin{itemize}[leftmargin=*]
    \item Introducing a pipeline (Section \ref{sub:derec}) to transform many-to-many collaborative data into one-to-many scenarios (Figure \ref{fig:common_data_structure}) for data synthesis
    \item Proposing the usage of conditional distribution (Section \ref{sub:conditional_distribution_indicator}) and similarity metrics (Section \ref{sub:improvement_counts} and \ref{sub:probabilistic_distance}) to measure individual column synthetic data fidelity 
    \item Developing an iterative algorithm to manage synthetic table fidelity in entirety (Section \ref{subsub:statistical-similarity}) using 'distribution of distribution similarity' (Figure \ref{fig:conditional_kernel})
\end{itemize}

\section{Related Works}
\label{sec:related-works}
We refer to papers regarding data synthesis, where limited amount of work on multi-table cases leads to overlooking of certain details that significantly impact synthetic data quality. 
\subsubsection*{\textbf{Privacy leakage when synthesising data. }} 
Research indicates that using transformers in data synthesis can enhance data security. While synthetic data aims to mitigate privacy leakage, studies show potential risks, particularly in ID-based data matching scenarios common in Data Clean Room \cite{synthetic-data-info-leak} \cite{herbrich2022data}. The \texttt{SiloFuse} framework \cite{shankar2024silofuse} offers a privacy-protecting solution by using distributed transformers to convert data into latent representations, making it impossible to retrieve the original data without access to the transformer. Consequently, language model-based synthesizers, particularly those incorporating multi-table synthesizers with LLM backbones (Section \ref{sub:derec-realtabformer}), are explored as a potential solution for enhancing synthetic data security. 
\subsubsection*{\textbf{Multi-table synthesizer architecture for data collaboration scenarios. }}
Multi-table synthesizers are designed for tables with a one-to-many relationship, as illustrated as the parent-child structure in \cite{SDV} \cite{solatorio2023realtabformer} \cite{shankar2024silofuse}. This architecture is often impractical in Data Clean Rooms, where formats like logbooks are frequently shared among users (Section \ref{sub:cross-table_management}). Despite significant advancements in single-table synthesizers like \texttt{CT-GAN} \cite{ctgan} and \texttt{TabDDPM} \cite{tabddpm}, these models struggle in data collaboration contexts (Figure \ref{fig:p-val-kernels-conditional}), highlighting the need for multi-table synthesizers (Table \ref{table:compare}). 
\subsubsection*{\textbf{Intuitive evaluation results for immediate performance monitoring. }} 
\cite{sdmetrics} and \cite{shankar2024silofuse} assess synthetic data quality using robust mathematical formulas and provide score-based outputs. However, while plotting these results for comparison, we found that the metric scores are biased in certain scenarios (Section \ref{sub:distribution_smoothing_bias}), highlighting the need for new metrics to address this bias. The Benjamini-Hochberg procedure \cite{benjamini-hochberg}, commonly used in medical research, offers a generalized performance analysis across various factors. This inspired the development of the \texttt{SIMPRO} evaluation metrics, which provide a comprehensive analysis and utilize illustrative plots for immediate performance comparison. 
\begin{table}[H]
  \centering
  \caption{\texttt{SiloFuse} provides privacy protection through tokenization but is limited to one-to-one data relationships. While \texttt{REaLTabFormer} supports one-to-many data synthesis with tokenization, the one-to-many assumption falls short in various collaborative data scenarios (Section \ref{sub:cross-table_management}).}
  \label{table:compare}
  \begin{tabular}{|c|c|c|c|}
    \hline
    \textbf{Tabular Data} & \textbf{Singular} & \textbf{Multiple} & \textbf{Token-} \\ 
    \textbf{Generator} & \textbf{Table} & \textbf{Table} & \textbf{-ization} \\ \hline
    \texttt{CT-GAN} \cite{ctgan} & \Checkmark & $\times$  & $\times$  \\ \hline
    \texttt{TabDDPM} \cite{tabddpm} & \Checkmark & $\times$  & $\times$ \\ \hline
    \texttt{SiloFuse} \cite{shankar2024silofuse} & \Checkmark & $\times$ & \Checkmark \\ \hline
    \texttt{REaLTabFormer}  \cite{solatorio2023realtabformer}& \Checkmark & \bcancel{\Checkmark} & \Checkmark \\ \hline
    \texttt{DEREC-REaLTabFormer} & \Checkmark & \Checkmark & \Checkmark\\ 
    (This work) & & & \\ \hline
  \end{tabular}
\end{table}

\section{Issues of multi-table synthesizers in realistic data collaboration scenarios}
There are two concerns on multi-table synthesis with real life data, namely privacy and architecture, and one concern on multi-table synthesis evaluation, that is evaluation metrics intuitiveness. Section \ref{sec:related-works} shows that the privacy concern can be well addressed by using language model-based synthesizer \cite{shankar2024silofuse}, such as the \texttt{REaLTabFormer} \cite{solatorio2023realtabformer}. This work then focuses on the architectural issue of multi-table synthesizers and the intuitiveness of evaluation metrics. 

\subsection{Multi-table Synthesizer Architecture}
\label{sub:multi-table-synthesizer-architecture}
\subsubsection*{\textbf{Cross-Table Management. }}
\label{sub:cross-table_management}
Multi-table synthesizers are not designed to manage across tables without pre-defined hierarchy. In Data Clean Room, tables from two parties do not necessarily have a parent table that contains one observation for each unique identifier (Table \ref{fig:unique-identifier}). Formal definitions are in Appendix \ref{appendix:formal-def}. Figure \ref{fig:unique-identifier} serves as an ideal parent table while Figure \ref{fig:repeating-identifier} acts as a child table. When both parties' data contain repeating identifiers (Table \ref{fig:repeating-identifier}), the multi-table synthesizer struggles to produce quality synthetic data without proper pre-processing (Figure \ref{fig:p-val-kernels-conditional}). To address this architectural issue, we propose the \texttt{DEREC} pre-processing pipeline, which creates a table with unique subjects to restructure the data for better compatibility with multi-table synthesizers. 
\begin{figure}[H]
    \begin{minipage}{.45\linewidth}
        \caption{Unique subject}
        \label{fig:unique-identifier}
        \centering
        \includegraphics[width = 0.7 \linewidth]{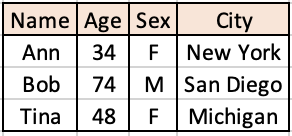}
    \end{minipage}%
    \begin{minipage}{0.55\linewidth}
        \caption{Repeating subjects}
        \label{fig:repeating-identifier}
        \centering
        \includegraphics[width = 0.9\linewidth]{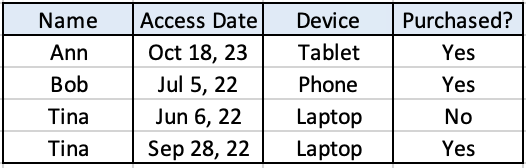}
    \end{minipage}
    \caption*{Different tabular structure in real life data. For multi-table synthesizer, Figure \ref{fig:unique-identifier} shows the required form for a parent table (Definition \ref{def:parent-table}).}
    \label{fig:different-structure}
\end{figure}


\subsubsection*{\textbf{Contextual Variation Disturbance. }} \label{sub:contextual_variation_disturbance} 
\textit{Contextual information} refers to data that remains constant within a sequence, such as a person's gender or age group (Section \ref{fig:common_data_structure}). However, this consistency may not always be observed, e.g., a person might use a friend's membership card, resulting in varying contextual columns like gender or purchase habits. This variation is particularly problematic when the frequency of subject occurrences is imbalanced, causing certain categories to dominate. This disruption indicates the need to separate contextual and non-contextual columns. To address this, \texttt{DEREC} (Section \ref{sub:derec}) extracts contextual columns from non-contextual ones to more accurately reflect the distribution of both types. Formal definition is in Appendix \ref{def:contextual-variable}. 

\subsubsection*{\textit{Example of Contextual Variation Disturbance.}} This work's experiments observe a case involving the age group column, which is expected to be contextual. Due to the high occurrence of a specific individual (with the same identifier), synthesizers misinterpret the distribution, leading to a dominant synthesis of that individual's age group, even among different synthesized individuals.

\subsection{Evaluation Metrics Intuitiveness}
\subsubsection*{\textbf{Conditional Consistency Focus. }}\label{sub:conditional_consistency_focus} 
Existing synthetic data evaluation metrics, like \texttt{SDMetrics} \cite{sdmetrics}, assume a parent-child table relationship, which is unsuitable for data collaboration due to the absence of a true hierarchy. For instance, \texttt{SDMetrics} evaluates multi-table synthetic data by comparing the cardinality of child rows in synthetic and original data. Instead, a comprehensive assessment of relationships across all columns is needed. Thus, the \texttt{SIMPRO} evaluation metrics are designed to measure these relationships in their entirety (Section \ref{sub:conditional_distribution_indicator}).  

\subsubsection*{\textbf{Distribution Smoothing Bias. }}\label{sub:distribution_smoothing_bias} Existing evaluation methods use mean-squared error function \cite{sdmetrics}, which favors smoothening the distribution shape. This distribution smoothening may not actually lead to improved synthetic data fidelity, since specific patterns may be neglected in the smoothened distribution. 
\subsubsection*{\textit{Example of Distribution Smoothing Bias. }} 
\begin{figure}[H]
    \begin{minipage}{.45\linewidth}
        \centering
        \includegraphics[width=0.8\linewidth]{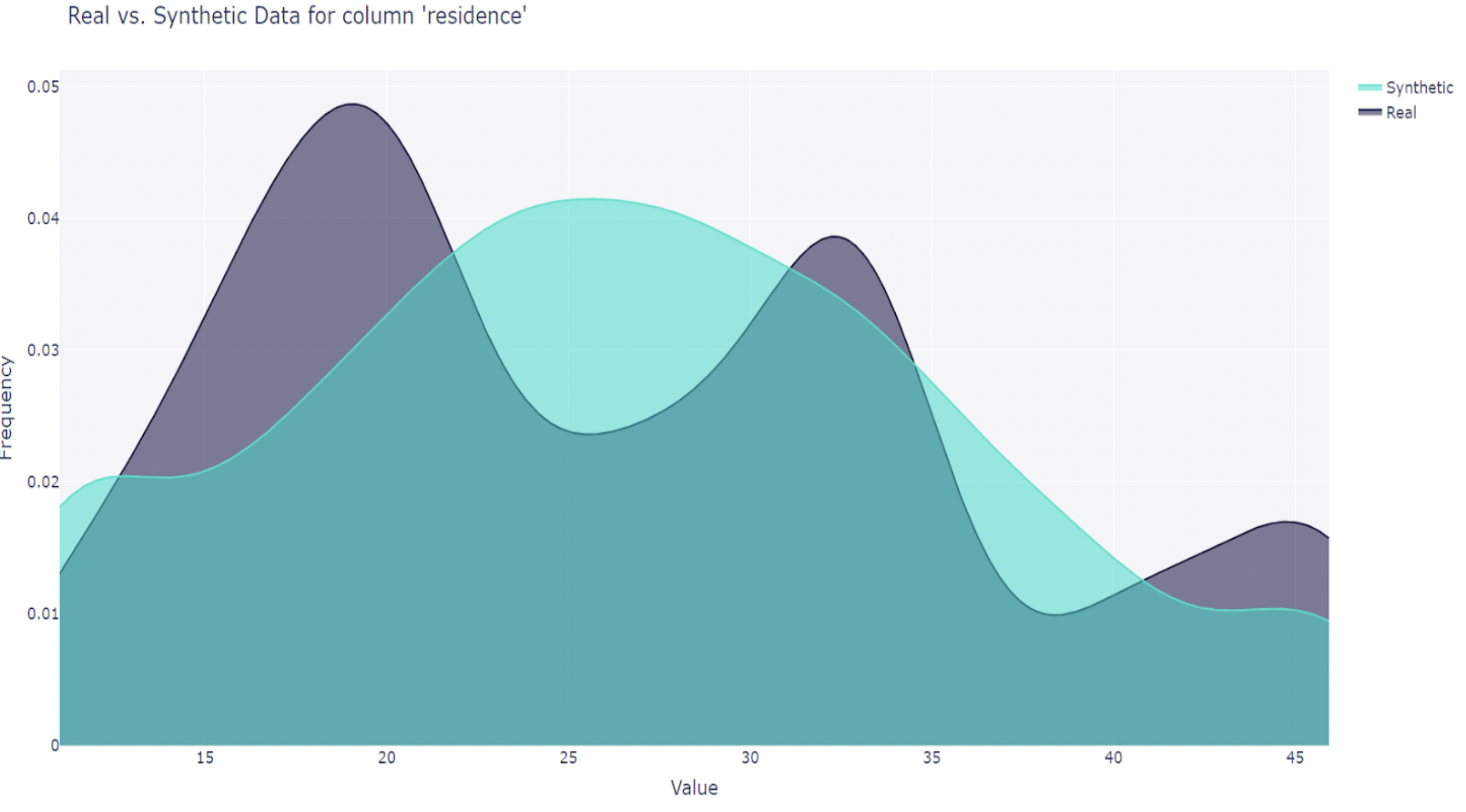}
        \caption*{\texttt{SDV} Column Shape Score: 84.79\%}
        \label{fig:sdv_hist}
    \end{minipage}%
    \begin{minipage}{0.55\linewidth}
        \centering
        \includegraphics[width=0.8\linewidth]{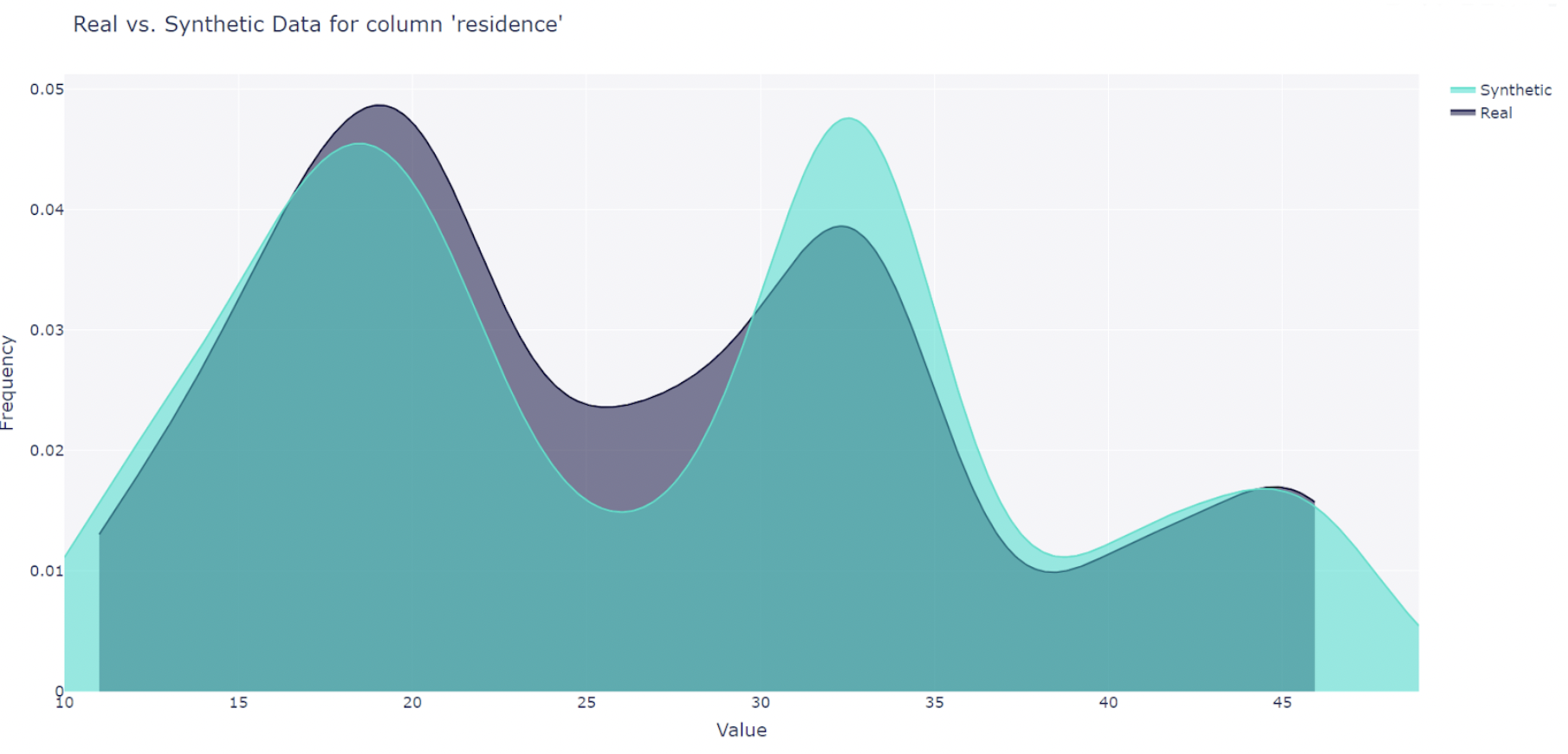}
        \caption*{\texttt{REaLTabFormer} Column Shape Score: 62.95\%}
        \label{fig:rtb_hist}
    \end{minipage}
    \caption{Histogram of synthetic data from \texttt{REaLTabFormer} shows a more similar distribution shape as the original data than \texttt{SDV}-synthesized data but is rated with lower score by the evaluation metrics (Section \ref{sub:distribution_smoothing_bias}), which implies curve smoothening may not facilitate capturing true distribution shape.}
    \label{fig:example_distribution_smoothing_bias}
\end{figure}

\texttt{SDMetrics} is used to evaluate synthetic data quality of \texttt{Synthetic Data Vault} ("\texttt{SDV}") and \texttt{REaLTabFormer}. In Figure \ref{fig:example_distribution_smoothing_bias}, the \texttt{SDMetrics} rates the \texttt{SDV}-synthesised data considerably higher than the \texttt{REaLTabFormer} data, but histograms show that the \texttt{REaLTabFormer} (in Figure \ref{fig:example_distribution_smoothing_bias}) can capture the distribution shape better.

\section{The \texttt{DEREC-SIMPRO} Framework}
The \texttt{DEREC-SIMPRO} framework combines the \texttt{DEREC} three-step pre-processing pipeline with the \texttt{SIMPRO} three-aspect evaluation metrics (Figure \ref{fig:overview}). The \texttt{DEREC} process involves (1) detecting contextual columns (formally defined in Appendix \ref{def:contextual-variable}), (2) recreating a parent table by extracting a single observation for each repeating subject, and (3) connecting this parent table with the remaining columns in a parent-child structure for multi-table synthesizers. Further details are provided in Sections \ref{subsub:detect}, \ref{subsub:recreate}, and \ref{subsub:connect}. The \texttt{SIMPRO} metrics evaluate synthetic data performance based on (1) statistical similarity to original data, (2) improvement counts of synthetic column pairs, and (3) probabilistic distance between original and synthetic data for a broader evaluation perspective. More details will be discussed in Sections \ref{subsub:statistical-similarity}, \ref{sub:improvement_counts}, and \ref{sub:probabilistic_distance}.

\subsection{\textbf{DEREC 3-steps Pre-Processing Pipeline}}
\label{sub:derec}
\begin{figure}[H]
\centering
\includegraphics[width = 1 \linewidth]{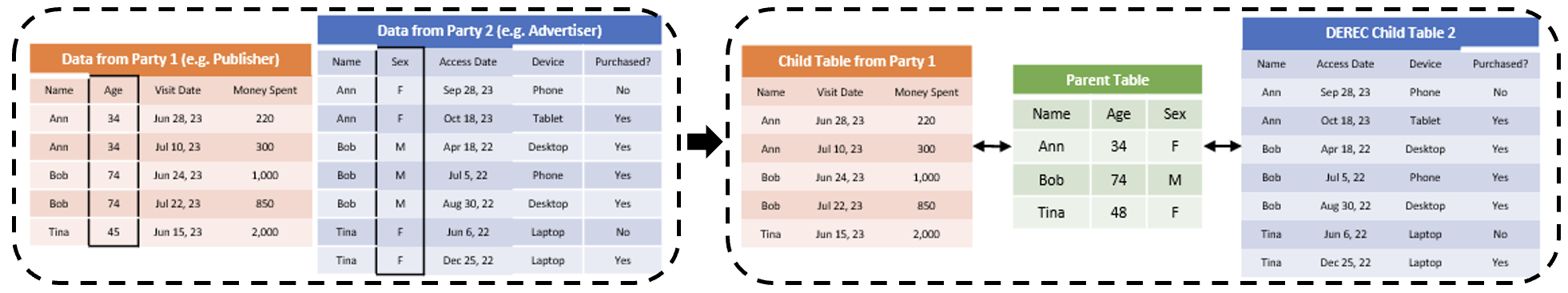}
\caption{Common multi-table structures before data synthesis (left subgroup) would be transformed into the parent-child structure (right subgroup) through the three-steps procedure (Section \ref{subsub:detect}, \ref{subsub:recreate}, \ref{subsub:connect}).}
\label{fig:common_data_structure}
\end{figure}

\texttt{DEREC} is designed to ensure up-to-standard performance of multi-table synthesizers, regardless of input tabular structure, addressing the common occurrence of tables with repeating subjects. The three steps, namely Detect, Recreate, and Connect, transform incompatible tables (left of Figure \ref{fig:common_data_structure}) into a compatible format for multi-table synthesizers. Each step will be discussed in detail, with accompanying algorithm examples provided in Appendix \ref{app:code}.

\subsubsection{\textbf{DEtect. }}
\label{subsub:detect}
The Detect step identifies contextual columns (Section \ref{sub:cross-table_management}) and mitigates \textit{Contextual Variation Disturbance}. This work shows the importance of addressing contextual variables during multi-tabular data synthesis (Section \ref{sub:contextual_variation_disturbance}), echoing a point made in the \texttt{SDV} work \cite{SDV} about the need for extra attention to contextual properties. With potential error in the dataset, it is impractical to have a threshold of 100\%, i.e. classifying the column to be contextual only if 100\% of unique subjects contain contextual information. For practicality, this study uses an arbitrary threshold of $M = 95\%$.

\subsubsection{\textbf{REcreate. }}
    \label{subsub:recreate}
In the Recreate step, an arbitrary parent table is formed using the contextual columns to tackle the \textit{Cross-table Management} challenge (Section \ref{sub:cross-table_management}). Since these columns contain repeating values for the same subject, retaining one observation per subject does not result in information loss. This table now has unique observations for each subject. This visualization of true contextual distribution helps mitigate the impact of \textit{Contextual Variation Disturbance} (Section \ref{sub:contextual_variation_disturbance}).

    \subsubsection{\textbf{Connect. }}
    \label{subsub:connect}
     This step connects the arbitrary parent table with the remaining columns as a parent-child pair. Remaining non-contextual columns are treated as the child tables. The multi-table synthesizer architecture (Section \ref{sub:multi-table-synthesizer-architecture}) is now satisfied. If there are non-contextual columns in both original datasets, they are considered as different child tables for two separate data synthesis rounds. Figure \ref{fig:common_data_structure} treats the remaining columns in the orange and blue tables as two separate child tables. These two child tables are then connected with the green parent table for separate data synthesis.

\subsection{\textbf{SIMPRO 3-aspects Evaluation Metrics}}
\label{sub:simpro}
 The \texttt{SIMPRO} evaluation metrics seek to provide simple and intuitive evaluation on synthetic data fidelity. It is based on the individual and overall statistical behaviour of the data. \texttt{SIMPRO} evaluates how well the \textit{cross-table feature correlation} is captured in three different aspects. The cross-table feature correlation is based on the conditional distribution of one column given the other column. 

\subsubsection*{\textbf{Conditional Distribution Indicator. }}
  \label{sub:conditional_distribution_indicator}
The conditional distribution indicator measures the similarity between synthetic and original data based on the conditional relationships between two columns. Switching the column order (e.g., Column A given B vs. Column B given A) allows for a more thorough analysis of causality on each feature. This similarity quantification enables apples-to-apples comparison between different synthesizers. This work proposes using the Kolmogorov-Smirnov Test for Goodness of Fit (KS-Test) \cite{kstest} and Wasserstein Distance (W-Distance) \cite{satone2021fund2vec} \cite{wdis2}, as both metrics have no pre-specified distribution assumptions \cite{ramdas2015wasserstein}. Detailed calculation algorithms are provided in Appendix \ref{algo:conditional_distribution_indicator}. After defining the fundamental cross-table feature correlation, three aspects to measure synthetic data are derived accordingly. 
\begin{itemize}
\item \textbf{S}tatistical similarity (Section \ref{subsub:statistical-similarity}) between original and synthetic data distribution to measure overall table similarity, pursuing for consistency in entirety (Section \ref{sub:conditional_consistency_focus}), 
    \item \textbf{IM}provement counts (Section \ref{sub:improvement_counts}) of synthesised column pairs fidelity to evaluate individual column similarity, addressing \textit{Distribution Smoothing Bias} (Section \ref{sub:distribution_smoothing_bias}), and
    \item \textbf{PRO}babilistic distance (Section \ref{sub:probabilistic_distance}) between original data and synthetic data for additional evaluation angle
\end{itemize}
\subsubsection{\textbf{Statistical Similarity. }}
\label{subsub:statistical-similarity}
The fidelity of synthetic data from different synthesizers is compared as a whole, evaluating both marginal distributions for each column and conditional distributions for column combinations. This aspect measures the similarity of these distributions between original and synthetic data \cite{mendelevitch2021fidelity}, drawing a "distribution of distribution similarity" for generalized comparisons across synthesizers. Distribution similarity can be assessed using goodness of fit test. This work focuses on using the KS-Test \cite{kstest} for its non-parametric property and sensitivity to central of distribution \cite{ks-test-2}. A large p-value indicates that we do not reject the hypothesis that the distribution of data generated by synthesizer \textbf{A} ($F_{Z;\textbf{A}}$) is similar to that from synthesizer \textbf{B} ($F_{Z;\textbf{B}}$), and vice versa. This is formally specified as a hypothesis test:
\begin{equation}\label{eq:hypothesis_testing_conditional_distribution_evaluation}
\begin{aligned}
    H_{0}: F_{Z;\textbf{A}}=F_{Z;\textbf{B}} \text{~~~versus~~~} H_{1}: F_{Z;\textbf{A}}\neq F_{Z;\textbf{B}}
\end{aligned}
\end{equation}
\subsubsection{\textbf{IMprovement Counts. }}
\label{sub:improvement_counts}
This aspect examines the performance of each cross-table feature correlation individually. Synthetic data fidelity is estimated by comparing the p-values from the KS-test statistic. If the KS-test p-value for a column from original data and synthesizer \textbf{A} is greater than that for synthesizer \textbf{B}, then \textbf{A} demonstrates higher fidelity for that column. Given that the difference of p-values are bounded to $[-1, 1]$ and are sensitive to noise \cite{GARCIA2015108}, a threshold is introduced to determine improvement. We denote the p-value-based cross-table feature correlation for column 2 given column 1 as $Q_{x^{2}|x^{1};\mathcal{P}}$ and set the threshold $T = 0.333$.  This categorizes results into "better", "no change", or "worsened", each occupying one-third of the p-value range. 
\begin{equation*}
    Q_{x^{2}|x^{1};\mathcal{P}}=\begin{Bmatrix}
        \text{better} & \text{if }\mathcal{P}_{x^{2}|x^{1};\textbf{A}} - \mathcal{P}_{x^{2}|x^{1};\textbf{B}}>T = 0.333\\
        \text{worsened} & \text{if }\mathcal{P}_{x^{2}|x^{1};\textbf{A}} - \mathcal{P}_{x^{2}|x^{1};\textbf{B}}<-T = -0.333\\
        \text{no change} & \text{otherwise}
    \end{Bmatrix}
\end{equation*}
 The numbers of improved cross-table feature correlation $N_{Q_{\mathcal{P}}=\text{improved}}$ is counted and compared with the numbers of worsened pairs $N_{Q_{\mathcal{P}}=\text{worsened}}$ to identify if there is net improvement \cite{count1} \cite{count2}. 

    \subsubsection{\textbf{PRObabilistic Distance. }}
    \label{sub:probabilistic_distance}
    In addition to p-values, distance metrics provide a useful measure of similarity \cite{Desai2021OnRO}. This work employs the Wasserstein Distance (W-Distance) \cite{wdis} \cite{wdis2} \cite{wdis3} to assess the distance between the conditional distributions of synthetic and original columns. A smaller W-Distance indicates higher similarity. We denote the distance-based cross-table feature correlation for column 2 given column 1 as $Q_{x^{2}|x^{1};\mathcal{W}}$ and set the threshold $T$ as the median ($\mathcal{M}$) of all correlations. This approach accounts for the unbounded nature of W-Distance differences, ensuring that a median difference is substantial enough to reduce noise effects.
    \begin{equation*}
        Q_{x^{2}|x^{1};\mathcal{W}}=\begin{Bmatrix}
        \text{better} & \text{if }\mathcal{W}_{x^{2}|x^{1};\textbf{A}} - \mathcal{W}_{x^{2}|x^{1};\textbf{B}}<-T=-\mathcal{M}\\
        \text{worsened} & \text{if }\mathcal{W}_{x^{2}|x^{1};\textbf{A}} - \mathcal{W}_{x^{2}|x^{1};\textbf{B}}>T=\mathcal{M}\\
        \text{no change} & \text{otherwise}
    \end{Bmatrix}
    \end{equation*} 

\section{Experimental Settings}
\subsection{Datasets and Assumptions}
\subsubsection*{Dataset. } 
The CTR Prediction - 2022 DIGIX Global AI Challenge \cite{digix_global_ai_challenge_2022} is used in this study. It contains two datasets, namely Advertisement and Feeds, both containing repeating 'user\_ID's. Due to the large data size, the data is subgrouped based on the type of advertisement tasks, resulting in eight task subgroups with over 750 observations each. All data, results, and related materials are stored in Google Drive, as the dataset size exceeds GitHub's 1GB repository limit. The link is as follows: \url{https://drive.google.com/file/d/1tZotBaeCkX0KypbmH-OAlo6LrHGYOW8w/view?usp=sharing}.
\subsubsection*{Assumptions. }Additional assumptions include: (1) no missing data, (2) only retaining subjects that co-exist in both tables (no orphan data), and (3) removing any non-classifiable string features. 

\subsection{\texttt{DEREC-REaLTabFormer}: This work}
\label{sub:derec-realtabformer}
 The \texttt{DEREC-REaLTabFormer} integrates \texttt{DEREC} (Section \ref{sub:derec}) with the \texttt{REaLTabFormer} synthesizer \cite{solatorio2023realtabformer}. The \texttt{REaLTabFormer} contains a \texttt{GPT-2} backbone for data generation, along with a quantile difference statistic measure to prevent overfitting and 'data copying' \cite{detect_data_copying}. \texttt{DEREC} addresses the remaining architectural challenges of multi-table synthesizers as discussed in Section \ref{sub:multi-table-synthesizer-architecture}. 
 \subsection{Comparable Models}
 \label{sub:comparables}
 The Control Group simulates a scenario where the parent-child requirement (Section \ref{sub:cross-table_management}) is ignored. The Feeds table with more contextual variables is chosen as the parent input. For benchmarking, single-table synthesizers \texttt{CT-GAN} and \texttt{TabDDPM} are used, with two datasets flattened into one by joining each subject with every possible combination. These benchmarks serve to show the need for multi-table synthesizers in data collaboration.


\section{Experimental Results using \texttt{SIMPRO} 3-aspects Evaluation Metrics}
\subsection{Statistical Similarity}
\label{sub:statistical-similarity-result}
\subsubsection*{\textbf{Different synthetic data quality from different models.}}
\begin{figure}[H]
    \centering
    \includegraphics[width = 0.95 \linewidth]{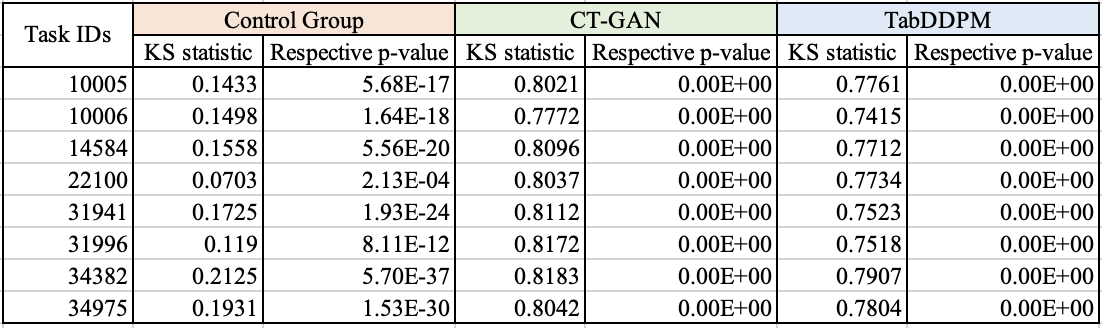}
    \caption{The Statistical Similarity evaluation metrics (Section \ref{subsub:statistical-similarity}) reveal significant differences in overall synthetic data distribution compared to the three baselines}
    \label{fig:ks_test}
\end{figure}

 The \texttt{DEREC-REaLTabFormer} generates synthetic data of significantly different quality compared to the three models discussed in Section \ref{sub:comparables}. By calculating the series of cross-table feature correlation (Section \ref{sub:conditional_distribution_indicator}), we assess 'distribution of distribution similarity' between original data and synthetic data generated by each synthesizer. Figure \ref{fig:ks_test} presents the p-values for the statistical similarity of these distributions. While all p-values are small, indicating low similarity, the Control Group shows a non-zero similarity, unlike the single-table synthesizers, indicating higher similarity between \texttt{DEREC-REaLTabFormer} and the Control Group. 

\subsection{Improvement Counts}
\label{sub:improvement-counts-results}
\subsubsection*{\textbf{Necessity of multi-table synthesizer in data collaboration. }}
Multi-table synthesizers consistently outperform single-table synthesizers. Table \ref{table: average_performance} highlights a significant improvement in cross-table feature correlation (Section \ref{sub:conditional_distribution_indicator}), while most values in the Control Group remain unchanged, contributing to a net improvement. This aligns with the results in Section \ref{sub:statistical-similarity-result}, where the difference between \texttt{DEREC-REaLTabFormer} and the Control Group is less pronounced than that between \texttt{DEREC-REaLTabFormer} and the single-table synthesizers. This supports the need for multi-table synthesizers in Data Clean Room. The next step involves a detailed comparison of performance with and without \texttt{DEREC} implementation.
\begin{table}[H]
    \centering
  \begin{tabular}{|c|c|c|c|}
    \hline
    \textbf{Performance} & \textbf{Control Group} & \textbf{CT-GAN} & \textbf{TabDDPM} \\ \hline
    Improved & 147 & 1362 & 1506 \\ \hline
    No Change & 1670 & 423 & 308 \\ \hline
    Worsened & 32 & 64 & 35 \\ \hline
  \end{tabular}
  \caption{The \texttt{DEREC-REaLTabFormer} outperformed single-table synthesizers dominantly and showed considerable net improvement against the Control Group.}
  \label{table: average_performance}
\end{table}
\subsubsection*{\textbf{Consistent net improvement after implementing \texttt{DEREC}}}
\begin{figure}
\centering
    \includegraphics[width=0.35\textwidth]{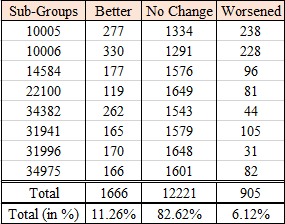} 
    \caption{The KS-test p-values indicate a consistent trend of more improved than worsened cross-table feature correlations, reflecting a net improvement in overall synthetic data fidelity after implementing the \texttt{DEREC} pipeline.}
    \label{fig:chi-squared-p}
\end{figure}

\texttt{DEREC} implementation brings a consistent net improvement in synthetic data fidelity (Figure \ref{fig:chi-squared-p}). Although the majority of column pairs show unchanged quality, there is a consistently greater number of pairs with improvements compared to those with worsened fidelity. This pattern indicates an overall net fidelity improvement. The results suggest the value of \texttt{DEREC}, by guaranteeing proper input requirement (Section \ref{sub:cross-table_management}) to optimize multi-table synthesizer performance.
\subsubsection*{\textbf{Improvement illustrations. }}
\label{sub:overall-table-improvement-illustrations}
\begin{figure}[t]
\centering
  \includegraphics[width=1\linewidth]{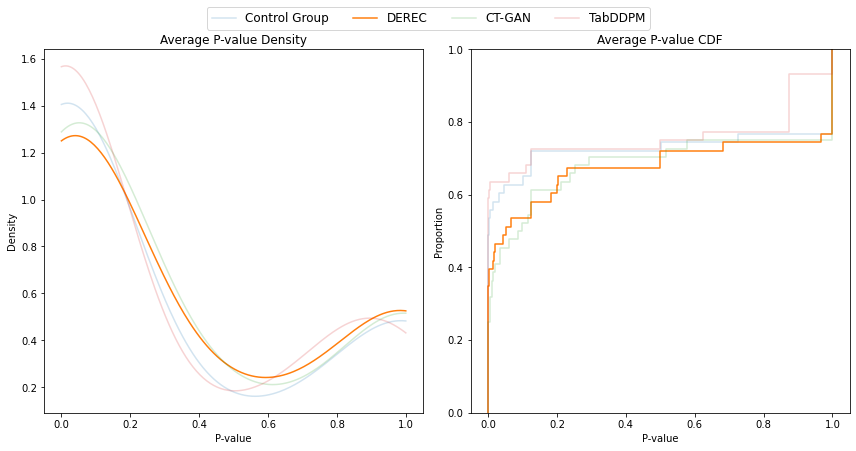}
  \caption{
  The shapes of the four models are fairly similar, with comparable densities at the endpoints, implying most models can understand individual column distribution well. }
  \label{fig:p-val-kernels-marginal}
\end{figure}

\begin{figure}[H]
      \centering
  \includegraphics[width=1\linewidth]{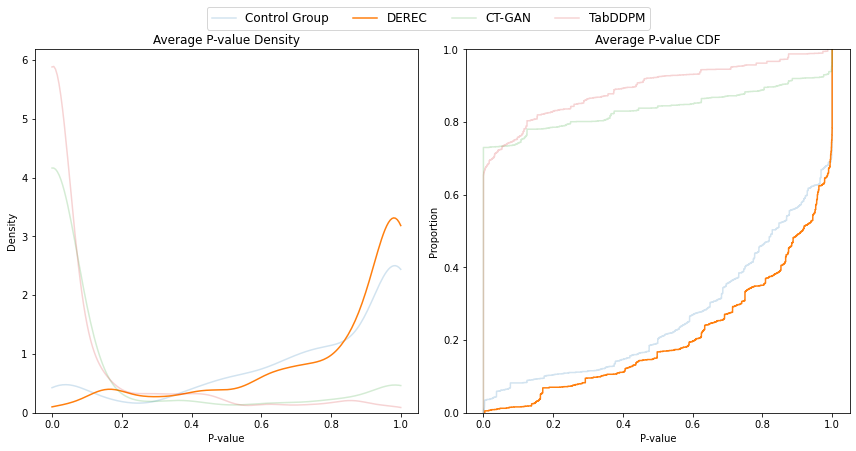}
  \caption{Single-table synthesizers perform significantly worse than multi-table synthesizers, regardless of being network-based as \texttt{CT-GAN} or diffusion-based as \texttt{TabDDPM}.}
  \label{fig:p-val-kernels-conditional}
\end{figure}

Inspired by the Benjamini-Hochberg Method \cite{benjamini-hochberg}, we plot both kernel and cumulative density graphs of the p-value-based cross-table feature correlations (Section \ref{sub:improvement_counts}). In the kernel density graph, a heavier right tail indicates more large p-values, implying higher similarity to the original data. For the cumulative density graph, better fidelity is inferred if the curve has a smaller area at low p-values and a sharp spike at larger p-values.\\ 
\\
In the one-dimensional marginal distribution case, Figure \ref{fig:p-val-kernels-marginal} shows that both \texttt{CT-GAN} and \texttt{DEREC-REaLTabFormer} capture the marginal distribution well, although the general shapes are similar across models. The \texttt{DEREC-REaLTabFormer} shows the lowest density at the lower p-values (0 to 0.2) but has a thicker density in the middle range (0.4 to 0.7), indicating slight outperformance. The \texttt{CT-GAN} also demonstrates improved performance, evident from a later spike in the cumulative density plot. This suggests that most models effectively model the individual distributions of each column. Individual subgroup plots are available in the Appendix (Figure \ref{fig:marginal_kernel}). \\
\\
In Figure \ref{fig:p-val-kernels-conditional}, both single-table synthesizers exhibit heavy tails at the lower end, indicating low similarity to the original data, hence their inadequacy for Data Clean Room. The shape of \texttt{DEREC-REaLTabFormer} closely mirrors that of the Control Group, but it demonstrates improvements across most ranges, with lighter densities in lower end, and a significantly heavier density in higher end, reinforcing the need to adhere to multi-table synthesizer input requirements. Individual subgroup plots are available in the Appendix (Figure \ref{fig:conditional_kernel}).

\subsection{Probabilistic Distance}
\label{sub:probabilistic-distance-results}
\subsubsection*{\textbf{Consistent results when using distance metrics. }}
\begin{figure}[H]
    \captionsetup{font=small}
    \centering
    \includegraphics[width=0.35\textwidth]{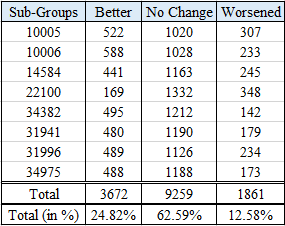} 
\caption{The W-Distance metrics results align with the p-value results from Section \ref{sub:improvement-counts-results}, confirming interchangeability across different metrics in measuring similarity.}
\label{fig:W-dis-performance}
\end{figure}
The W-Distance-based cross-table feature correlation serves as an additional metric to validate the results. Figure \ref{fig:W-dis-performance} shows a similar pattern to the p-value analysis, indicating consistent net improvement across all subgroups.

\subsection{Result Implication}
\label{sub:further-discussion}
Two areas for improvement in the \texttt{DEREC} pipeline are identified for future research. (1) \textbf{Incorporating Cross-Child-Table Feature Correlation}: The current \texttt{DEREC} pipeline (Section \ref{sub:derec}) does not fully capture all cross-table feature correlations, as it synthesizes the parent table independently with each child table. A unified training approach could enhance model performance by better capturing these correlations. (2) \textbf{Using Advanced Language Model Backbones}: While \texttt{REaLTabFormer} currently uses the \texttt{GPT-2} model, future implementations of advanced models like \texttt{GPT-4o} \cite{gpt4o} or \texttt{Llama 3} \cite{touvron2023llama} may yield better results. \\ 
\\
A coding error in the \texttt{REaLTabFormer} package is also found. During bootstrapping, a folder named 'not-best-disc-model' is created to store model weights with the second-best results. However, weights are only saved under a specific condition, which may leave the folder empty and cause the program to crash when it attempts to retrieve weights. This issue was resolved by adding code (details in Appendix \ref{sec:coding-error}) to ensure that current weights are saved in the folder if it is empty, with these weights later replaced by improved results.

\section{Conclusion} 
This work identifies a gap between multi-table synthesizer architecture and real-world data collaboration in Data Clean Room, proposing the \texttt{DEREC} 3-step pipeline to adapt real-life data to a parent-child structure (Section \ref{sub:cross-table_management}). This transformation helps mitigate contextual variable disturbances (Section \ref{sub:contextual_variation_disturbance}) by creating an arbitrary parent table that reflects true variable distributions. However, there are still opportunities to improve \texttt{DEREC} by including all cross-child-table feature correlations. Additionally, inefficiencies in current multi-table evaluation metrics are noted, particularly in their emphasis on parent-child correlations (Section \ref{sub:conditional_consistency_focus}) and preference for smoothed distributions (Section \ref{sub:distribution_smoothing_bias}). The \texttt{SIMPRO} 3-aspect evaluation metrics is designed to evaluate multi-table synthetic data both individually and collectively using cross-table feature correlations (Section \ref{sub:conditional_distribution_indicator}), effectively addressing these existing challenges.

\clearpage
\bibliographystyle{plain}
\nocite{*}
\bibliography{ref}

\clearpage

\appendix

\section{Formal Definition for Section \ref{sub:cross-table_management}} \label{appendix:formal-def}
\begin{definition}[Parent Table]
    \label{def:parent-table}  
Parent table $\textbf{T}$ is defined to be a table containing different observations with $n$ features such that for each unique identifier $i$, $o_{i} = [x_{i1}, x_{i2}, \dots, x_{in}]$. 
\end{definition}
\begin{definition}[Child Table]
    \label{def:child-table}
A child table \textbf{s} contains some observations on every unique parental observation $o_{i}$. For a child table subset that only contains observations from $o_{i}$, 
\begin{equation} \label{eq:parent_notation_simp}
    s_{o_{i}} = \begin{bmatrix}
    o^{(1)}_{i} \\ 
    o^{(2)}_{i} \\
    \cdots \\
    o^{(k)}_{i}
\end{bmatrix} = \begin{bmatrix}
    x'^{(1)}_{i1} & x'^{(1)}_{i2} & \cdots & x'^{(1)}_{im}\\
    x'^{(2)}_{i1} & x'^{(2)}_{i2} & \cdots & x'^{(2)}_{im}\\
    \cdots \\
    x'^{(k)}_{i1} & x'^{(k)}_{i2} & \cdots & x'^{(k)}_{im}
\end{bmatrix}
\end{equation}
\end{definition}
\section{Formal Definition for Section \ref{sub:contextual_variation_disturbance}} 
\label{def:contextual-variable} 
\begin{definition}[Contextual Information] 
 Rewrite $s$ from \eqref{eq:parent_notation_simp}:
    \begin{equation}\label{eq:parent_notation_detail}
    s = \begin{bmatrix}
    o^{(1)}_{1} \\
    o^{(2)}_{1} \\
    o^{(3)}_{1} \\
    \cdots \\
    o^{(1)}_{2} \\
    o^{(2)}_{2} \\
    o^{(3)}_{2} \\
    \cdots \\
    o^{(1)}_{l} \\
    \cdots
\end{bmatrix} = \begin{bmatrix}
    x'^{(1)}_{11} & x'^{(1)}_{12} & \cdots & x'^{(1)}_{1m} \\
    x'^{(2)}_{11} & x'^{(2)}_{12} & \cdots & x'^{(2)}_{1m} \\
    x'^{(3)}_{11} & x'^{(3)}_{12} & \cdots & x'^{(3)}_{1m} \\
    & \cdots \\
    x'^{(1)}_{21} & x'^{(1)}_{22} & \cdots & x'^{(1)}_{2m} \\
    x'^{(2)}_{21} & x'^{(2)}_{22} & \cdots & x'^{(2)}_{2m} \\
    x'^{(3)}_{21} & x'^{(3)}_{22} & \cdots & x'^{(3)}_{2m} \\
    & \cdots \\
    x'^{(1)}_{l1} & x'^{(1)}_{l2} & \cdots & x'^{(1)}_{lm} \\
    & \cdots 
\end{bmatrix}
\end{equation}

Consider the column $x'_{n}$
\begin{align*}
    x'_{n} &= \begin{bmatrix}
    x'^{(1)}_{1n} & x'^{(2)}_{1n} & 
    \cdots &
    x'^{(1)}_{2n} &
    x'^{(2)}_{2n} &
    \cdots &
    x'^{(1)}_{ln} &
    \cdots 
\end{bmatrix}^\top \\
& =\begin{bmatrix}
    \vec{x}'_{1n} &
    \vec{x}'_{2n} &
    \cdots &
    \vec{x}'_{ln} 
\end{bmatrix}^\top
\end{align*}
If $x'^{(a)}_{1n} = x'^{(b)}_{1n} \forall a \neq b$, then $\vec{x}'_{1n}$ is contextual for unique identifier $i = 1$. If over $M$\% $\vec{x}'_{dn}\forall d \in \{\text{unique identifier}\} \text{ is contextual }$, $x'_{n}$ is a \textit{contextual column}. 
\end{definition}

\section{Detailed Algorithm for Section \ref{sub:conditional_distribution_indicator}} 
\label{algo:conditional_distribution_indicator}
\begin{algorithm}[H]
\caption{Cross-table Feature Correlation computation}
\begin{enumerate}
    \item Given original columns 1 and 2 $\vec{x}_{1;O}, \vec{x}_{2;O}=\begin{pmatrix}
    x^{1;O}_{1} & x^{1;O}_{2} & x^{1;O}_{3} & \cdots
\end{pmatrix}^{\top}, \begin{pmatrix}
    x^{2;O}_{1} & x^{2;O}_{2} & x^{2;O}_{3} & \cdots
\end{pmatrix}^{\top}$, and synthetic columns 1 and 2 $\vec{x}_{1;syn}, \vec{x}_{2;syn} =\begin{pmatrix}
    x^{1;syn}_{1} & x^{1;syn}_{2} & x^{1;syn}_{3} & \cdots
\end{pmatrix}^{\top}$, $ \begin{pmatrix}
    x^{2;syn}_{1} & x^{2;syn}_{2} & x^{2;syn}_{3} & \cdots
\end{pmatrix}^{\top}$
    \item Compute $P(x^{2}_{j}|x^{1}_{i})$ for column 2 value $x^{2}_{j}$ given column 1 value $x^{1}_{i}$
    \item Repeat (2) for every $x^{2}_{j}$ to obtain the conditional distribution of $x^{2}|x^{1}_{i}$
    \item Implement (2) to (3) on both the original and synthetic data to obtain $x^{2;O}|x^{1;O}_{i}$ and $x^{2;syn}|x^{1;syn}_{i}$
    \item Measure the similarity of the distributions with different statistical tools to obtain a similarity indicator given column 1 value $x^{1}_{i}$. KS-Test (Section \ref{sub:improvement_counts}) and W-Distance (\ref{sub:probabilistic_distance}) are recommended. 
    \item Repeat (2) to (5) on every column 1 value $x^{1}_{i}$ to obtain a conditional distribution given every column 1 value $Z_{x^{2}|x^{1}_{i}} \forall x^{1}_{i}$
    \item Compute cross-table feature correlation $Z_{x^{2}|x^{1}}$ by taking the weighted average of all $Z_{x^{2}|x^{1}_{i}}$ with the probability of the occurrence for every unique parent value $P(x^{1}_{i})$
    \item Repeat on every column pair to obtain a series of cross-table feature correlation
\end{enumerate}
\end{algorithm} 

\section{Supplementary algorithm for \texttt{DEREC} pipeline in Section \ref{sub:derec}}
\label{app:code}
\subsection{Algorithm of the \texttt{DEREC} pipeline}
\begin{algorithmic}

    \If{number of unique identifier with only one category / number of unique identifier $= 0.95$}
        \State column $\gets$ contextual
    \Else
        \State column $\gets$ non-contextual
    \EndIf\\
    \textbf{Repeat} for all columns\\
    \textbf{Repeat} for all tables\\
    Parent Table $\gets$ all contextual columns\\
    Child Table $\gets$ all non-contextual columns
\end{algorithmic}

\subsection{Python Code Example}
\begin{lstlisting}
parent_col = []
child_col_table1 = []
child_col_table2 = []

for every column in table1:
    count = 0
    for every unique_ID:
        if column.nunique() == 1:
            count += 1
    if count / count(unique_ID) >= 0.95:
        parent_col.append(column)
    else:
        child_col_table1.append(column)

for every column in table2:
    count = 0
    for every unique_ID:
        if column.nunique() == 1:
            count += 1
    if count / count(unique_ID) >= 0.95:
        parent_col.append(column)
    else:
        child_col_table2.append(column)
        
return parent_col, child_col_table1, 
child_col_table2
\end{lstlisting}
\clearpage
\section{Supplementary results for Section \ref{sub:overall-table-improvement-illustrations}}
\begin{figure}[H]
    \centering
    \includegraphics[width=0.75\columnwidth]{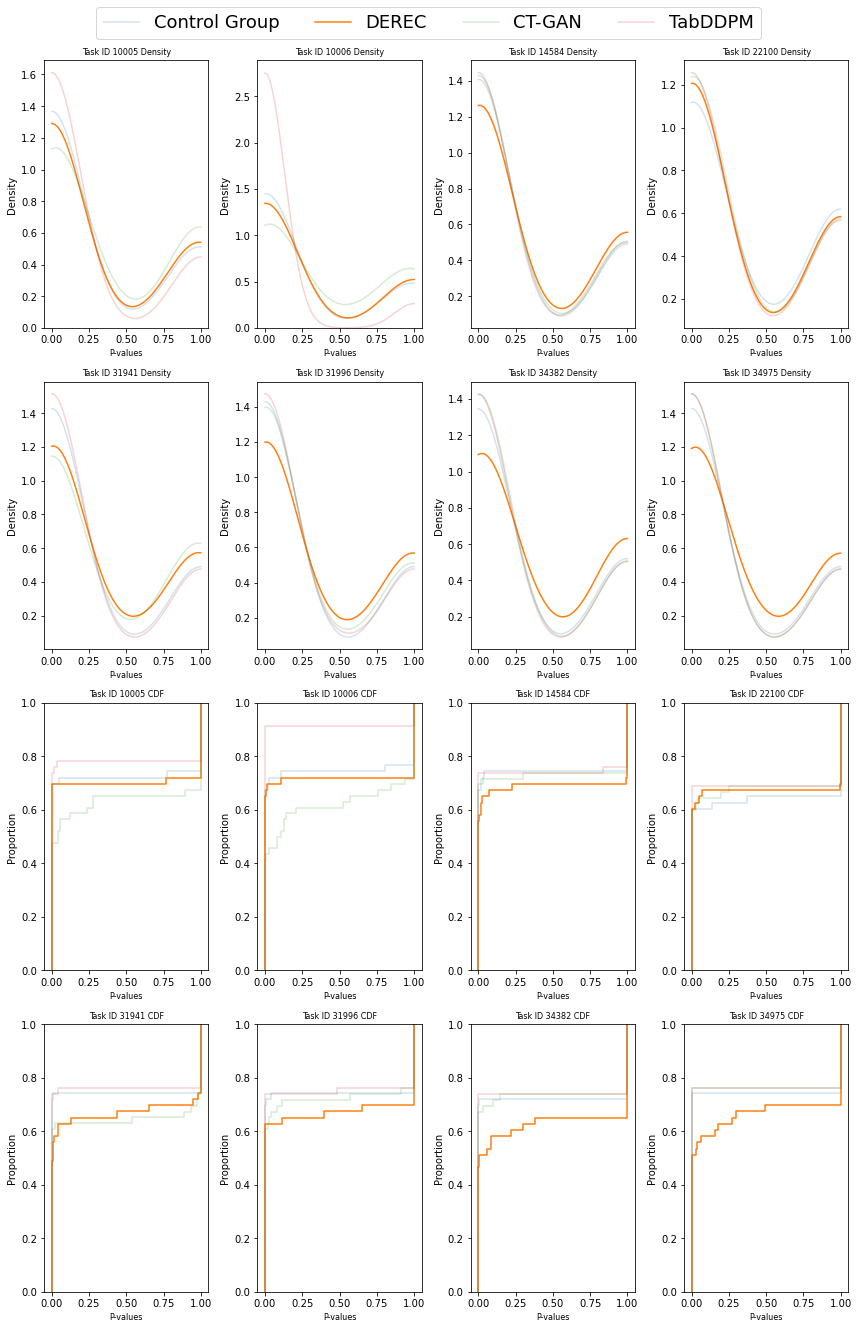}
    \caption{Density of Column Marginal Distribution in each task ID subgroup showing general outperformance of the \texttt{DEREC-REaLTabFormer} and \texttt{CT-GAN}}
    \label{fig:marginal_kernel}
\end{figure}

\begin{figure}[H]
    \centering
    \includegraphics[width=0.75\columnwidth]{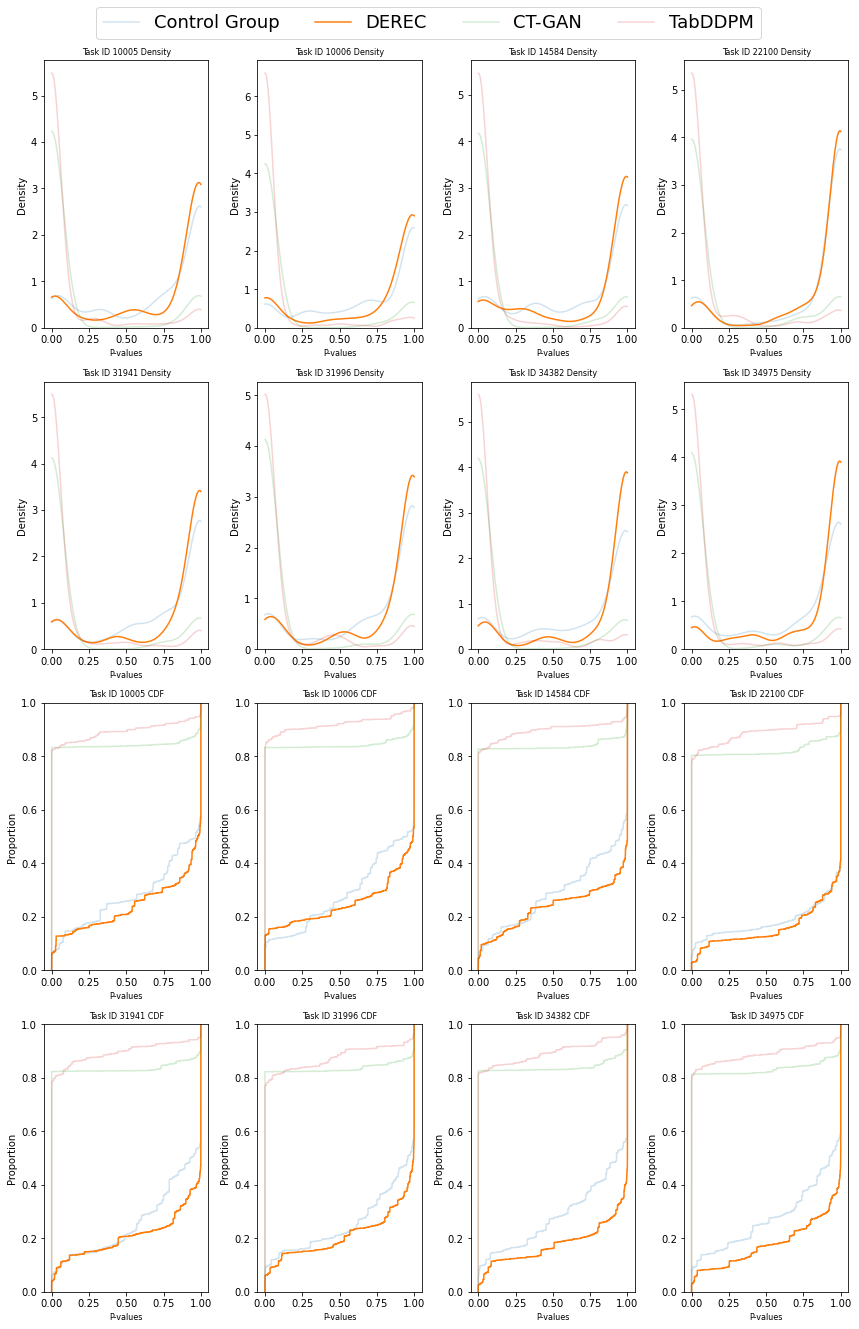}
    \caption{Density of Column Conditional Distribution in each task ID subgroup showing general outperformance of the \texttt{DEREC-REaLTabFormer}}
    \label{fig:conditional_kernel}
\end{figure}

Figures \ref{fig:marginal_kernel} and \ref{fig:conditional_kernel} supplementary the results for this work's claim on Figures \ref{fig:p-val-kernels-marginal} and \ref{fig:p-val-kernels-conditional}. 

\clearpage

\section{Coding Error of \texttt{REaLTabFormer} package}
\label{sec:coding-error}
The original code was written as in line 791 of the 'realtabformer.py' file in the \texttt{REaLTabFormer} package folder as follows:
\begin{figure}[H]
    \centering
    \includegraphics[width=0.8\linewidth]{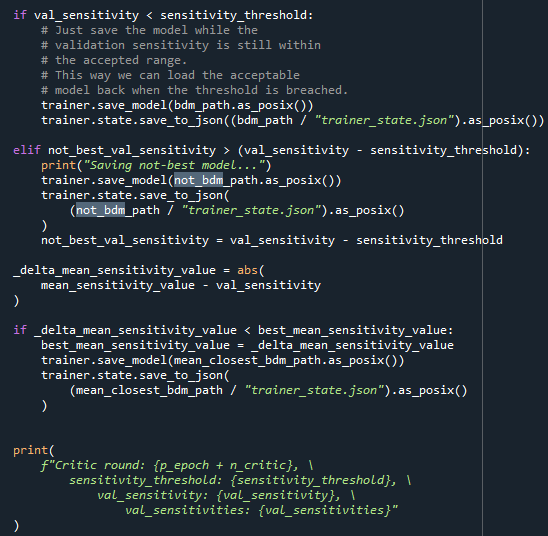}
\end{figure}
\textcolor{white}{}
\\
The 'not-best-model' named as 'not\_bdm\_path' in the code will only be saved in the else-if-statement. In case where the else-if-statement is not triggered, the 'not-best-model' folder will be empty, and subsequent weighting calls on the 'not-best-model' from the 'mean-best-model' (named as 'mean\_closest\_bdm' in the code) will crash the program. To solve that, the following code is added into the code that saves the 'mean-best-model' weightings: 
\begin{figure}[H]
    \centering
    \includegraphics[width=0.8\linewidth]{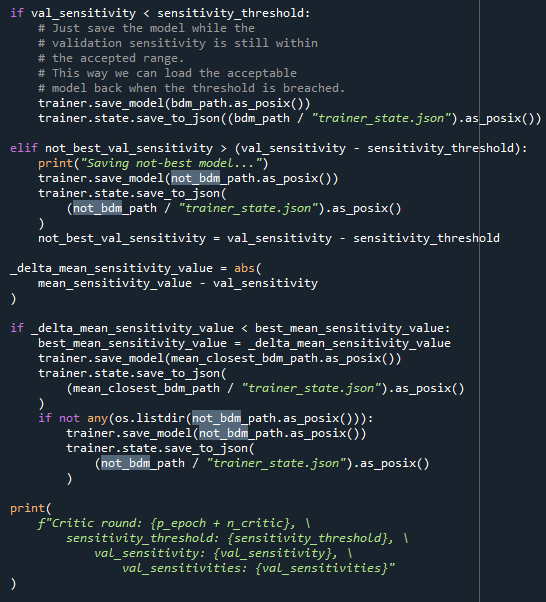}
\end{figure}
\textcolor{white}{blank}\\
\\
This ensures that the 'not-best-model' folder will never be empty if there is any saving in the 'mean-best-model', hence making sure there is always a valid object upon any weighting calls. 

\end{document}